# Evidence Gap Maps as Critical Information Communication Devices for Evidence-based Public Policy

Esteban Villa-Turek♠     Hernan David Insuasti-Ceballos♣

Jairo Andres Ruiz-Saenz♦     Jacobo Campo-Robledo*

*Abstract*— The public policy cycle requires increasingly the use of evidence by policy makers. Evidence Gap Maps (EGMs) are a relatively new methodology that helps identify, process, and visualize the vast amounts of studies representing a rich source of evidence for better policy making. This document performs a methodological review of EGMs and presents the development of a working integrated system that automates several critical steps of EGM creation by means of applied computational and statistical methods. Above all, the proposed system encompasses all major steps of EGM creation in one place, namely inclusion criteria determination, processing of information, analysis, and user-friendly communication of synthesized relevant evidence. This tool represents a critical milestone in the efforts of implementing cutting-edge computational methods in usable systems. The contribution of the document is two-fold. First, it presents the critical importance of EGMs in the public policy cycle; second, it justifies and explains the development of a usable tool that encompasses the methodological phases of creation of EGMs, while automating most time-consuming stages of the process. The overarching goal is the better and faster information communication to relevant actors like policy makers, thus promoting well-being through better and more efficient interventions based on more evidence-driven policy making.

**Keywords:** *information systems*, *evidence gap maps, efficient communication, evidence-based public policy, structured literature search, impact evaluation.*

♠ Northwestern University, Evanston. Contact: villaturek@u.northwestern.edu
♣ Departamento Nacional de Planeación, Colombia. Contact: hinsuasti@dnp.gov.co
♦ Departamento Nacional de Planeación, Colombia. Contact:: jairoruiz@dnp.gov.co
* Departamento Nacional de Planeación, Colombia. Contact: jacampo@dnp.gov.co

# 1 Introduction

Evidence-based public policy improves the quality of life of individuals and their relationship with the state, and as such it is a fundamental pillar in the efforts towards its modernization. This effort involves improving the state's ability to deliver on citizens' needs to increase their trust in public administration. The use of evidence in government decision-making has increased considerably regarding the use of evidence in the public policy cycle, which includes the formulation, design, implementation, and evaluation of public policies. However, this evidence can be presented in very different formats, such as scientific or academic articles, databases, reports, websites, among others. This poses the following challenges to policy makers: 1) to materially have access to the necessary evidence for the formulation, design, implementation, and evaluation of public policies; and 2) to assess if and where evidence gaps exist around certain topics of interest, including the impacts and results of related policy interventions.

Evidence Gap Maps (EGMs) have been promoted extensively by the International Initiative for Impact Evaluation (3ie) and were conceived as a response to this challenge and to the need to facilitate evidence-based decision making (B Snilstveit et al., 2013; Birte Snilstveit et al., 2016, 2017). In essence, they are instruments that allow for the mapping and synthesizing of large volumes of relevant evidence to offer a general vision of existing evidence about the impact and results of specific public policy interventions on larger outcomes of interest. The process, however, can take up to several months. This document offers a methodological review of EGMs and announces the development of a usable system that comprises all steps of EGM creation and automates some of them by connecting directly to available APIs of academic research databases to query and download articles that meet determined inclusion criteria. The user might be any stakeholder working in academia, civil society or in government and who actively engages in the process, as we will discuss below.

The contribution of the document is two-fold. First, it presents the critical importance of EGMs in the public policy cycle. Second, it justifies and explains the development of a usable tool that encompasses the methodological phases of creation of EGMs, while automating most time-consuming and resource expensive stages in the process. Additionally, this document contributes to the empirical literature on EGMs and serves as a support to readers and academics interested in evidence-based policy making.

This document is organized as follows. In the second section EGMs are introduced, emphasizing the uses and advantages they possess, and establishes the link between EGMs and the public policy cycle. A methodological review of EGMs is presented in the third section. The fourth section presents the justification for the development of a semi-automatic tool to support the creation of EGMs. Finally, the limitations and final considerations are presented in the fifth and sixth sections.

## 2 Evidence Gap Maps (EGMs) and Evidence-based Public Policy

As mentioned above, EGMs are instruments aimed at obtaining a general, intuitive, and timely view of the evidence or the lack thereof at certain intersections of public policy interventions and outcome variables. Their objective is to promote better decision-making in policy making or when making decisions to allocate research resources to generate evidence in areas where it is scarce, and their importance lies precisely in the fact that they allow an overview of the evidence-generating studies that exist around the world regarding public policy interventions and their corresponding results.

This possibility of having a quick and complete vision of what has been done and what has happened policy-wise, is therefore a fundamental element of evidence-based public policy, which strengthens its creation, implementation, and evaluation processes. This panorama is important, because it allows public policy makers to have a broad and comparative position when faced with the uncertainty inherent in the design of new interventions and thus, we argue, EGMs should be widely used when designing public policy interventions.

Public policies are by nature a forward-looking practice, destined to shape certain scenarios that we consider desirable. One of the main tools to correctly approach this is the use of evidence, understood as a set of systematic observations identified for the purpose of establishing facts and/or testing hypotheses, which were obtained through replicable methodologies (Nutley, 2003). In this way, evidence-based public policy should be understood as an approach that aims to ensure that decisions on public policy are guided by evidence, which can be generated through quantitative and qualitative methods. This use of evidence allows, firstly, to differentiate it from any other type of knowledge (e.g., expert opinions) and, secondly, it implies the documentation of research methods, peer review and public scrutiny, which favors the confidence in the results derived from the analysis of this type of evidence (Nutley et al., 2013). In general, greater use of evidence is associated with greater probabilities of achieving the social and economic objectives of a wide array of programs and projects proposed by a government, obtaining better results for the population, and saving valuable limited resources by selecting more effective or profitable solutions for social problems (Chalmers, 2003).

This essential aspect of evidence for policy making is therefore why we argue it is critical to gather as much of it as possible and to use it to justify the choice of certain public policy interventions. This is where EGMs become crucial, and they have been used in the past as very effective ways to synthesize the vast amounts of evidence there is to inform public policy decisions. Several recent examples can be seen in EGMs regarding the traditional, complementary and integrative medicines during COVID-19 (Portella et al., 2020); the support of local institutions for green growth (Berkhout et al., 2018); interventions for persons with disabilities in low- and middle-income countries and their effectiveness (Saran et al., 2020); interventions for reducing violence against children in low- and middle-income countries (Pundir et al., 2020); interventions against institutional child maltreatment (Finch et al., 2021); performance measures and management in primary healthcare systems in low- and middle-income countries (Munar et al., 2019).

## 3   Methodological review of EGMs

The methods used for constructing EGMs draw on methodologies previously implemented in other efforts of evidence mapping and synthesis approaches (Birte Snilstveit et al., 2016). Based upon them there are five main steps involved in designing and constructing EGMs, which are outlined below.

   i)   *Definition of Interventions and Outcomes Framework*

The first step in the EGM construction process requires previous knowledge and research on the topic being investigated. As such, usually a team of researchers conduct a preliminary literature review that allows them to assess existing interventions around the policy area of interest, as well as the outcome universe that might be affected by them (Birte Snilstveit et al., 2016). During this framework definition process policy researchers should generate communication channels to consult with interested actors and stakeholders, as they are crucial sources of knowledge on general relevance and acceptability of the proposed framework (B Snilstveit et al., 2013).

*ii) Determination of Inclusion Criteria*

Generally, inclusion criteria for studies in a new EGM stem directly from the framework definition step because the substantive and field-specific requirements are defined and set forth then, as outlined above. Nevertheless, there can be formal differences regarding the type of documents to include in the EGM depending on its primary purpose. If, for example, the primary goal of the EGM is to inform decision and policy makers regarding existing evidence available in relation to interventions of interest, it could be best to only include systematic reviews that best curate and synthetize a possibly large volume of primary studies. If, on the other hand, the objective of the EGM is to identify existing research gaps, the primary documents to process would necessarily be primary studies, like impact evaluations, which could more easily translate into a quantitative overview of the amounts of evidence being produced in any given intervention-outcome intersection, or the lack thereof (Birte Snilstveit et al., 2016), which could entail a valuable signal for research resource allocation decisions. This gap-related functionality of EGMs can, in turn, signal the existence of *absolute gaps* if there is little or no impact evaluations or primary studies in the intersection, and *synthesis gaps,* located in intersections where multiple impact evaluations or primary studies exist, but there is no or not recent synthesis of them, usually in the form of a systematic review (Birte Snilstveit et al., 2017).

*iii) Search and Inclusion Assessment*

Once the substantive and formal criteria for inclusion have been set, the next step is to search and screen for studies that fit both sets of requirements. To do it sustainably, a delicate balance of breadth and depth must be attained. More precisely, it has been established that highly sensitive searches with low precision is unmanageable, reason for which more basic yet systematic searches using keywords should be prioritized (Birte Snilstveit et al., 2016). Search methods necessarily must respond to the ultimate purpose of the EGM, as established in the second step. If the primary goal is to present a translated synthesis of existing evidence for decision and policy makers, the search must target mainly systematic reviews repositories and databases. Otherwise, the search methods should focus on reputable sources of primary studies like impact evaluations, which usually requires more in-depth searches and screenings (B Snilstveit et al., 2013). In either case, researchers should supplement said strategic search methods with studies from other reputable sources using, for instance, snowballing or citation tracking techniques (Birte Snilstveit et al., 2016).

iv) *Data Extraction, Coding and Critical Appraisal*

After having searched and obtained all relevant documents, the next step is to systematically extract all necessary information and code it in a structured way (Birte Snilstveit et al., 2016). Depending on the scope of the EGM and its goal, the coding can be tailored to reflect the occurrence of any given intervention-outcome intersection, as well as the inclusion of other plausibly relevant insights, such as methodologies implemented, status of the study, geographical scope, etc. (B Snilstveit et al., 2013). Finally, critical appraisals of systematic reviews, their quality ratings and user-friendly summaries of the most relevant documents identified can be included in the EGMs, for instance in their interactive visualization (B Snilstveit et al., 2013).

v) *Analysis and user-friendly presentation*

The last step is to populate the EGM itself, which comprises a descriptive and non-formal representation of previously extracted and coded information placed onto the intervention-outcome intersection framework matrix, as defined in the first step of the process (Birte Snilstveit et al., 2016). This allows for a comprehensive and visual overview of the state of the research in each intersection of interest, enabling researchers to find study gaps and even signaling the quality of reviewed systematic reviews using color codes, all of which can be filtered based on characteristics like geographic location, type of study, population, etc. (Birte Snilstveit et al., 2016). A good practice is to include an explanatory note in the EGM containing a description of the methodologies employed in its construction, and, if possible, brief summaries touching upon identified policy implications, future research, findings summaries, among others (Birte Snilstveit et al., 2016).

## 4  New Horizons: Applied Computational Methods for EGMs

We have outlined the general methodology proposed for the creation of EGMs. The five steps required for the successful design, construction and utilization of EGMs are an important step towards their standardization and generalized use but can imply major drawbacks. Specifically, it has been identified that normally it takes very large amounts of time to complete some of the EGM creation steps. Phases ii) and iii) where seemingly large volumes of documents must be searched for and screened, and information needs to be extracted and manually coded are especially prone to be notoriously taxing, both in terms of human and time resources, which can lead to EGMs taking sometimes up to 6 months to complete (Birte Snilstveit et al., 2016).

If EGMs are instruments intended for timely overviews of evidence or lack thereof aimed at supporting better decision making in rapidly changing policy processes or resource allocation calls, immediacy in their availability should be of the uttermost importance and requiring months towards their completion seems to play the opposite role. For this reason, we are proposing novel and efficient ways of applying computational methods and natural language processing (NLP) techniques that seek to automate as much as possible the most resource-intensive aspects of EGM construction.

In recent years, text-mining has become a salient priority for some of the major publishing companies specializing in academic, scientific, and technical literature, through the design and implementation of application programming interfaces (APIs). Indeed, text-mining has been used in other mapping exercises and can prove to be instrumental for more efficient creation of EGMs (Birte Snilstveit et al., 2016). Notable examples of the latter can be found in the noteworthy technical developments undertaken by salient publishers like Elsevier (SCOPUS & ScienceDirect), Springer Nature and CORE. All offer API implementations that let authenticated researchers establish a direct connection with their databases and servers to interact with them in a dynamic way, allowing them to send requests containing all relevant search query terms and receiving all matched documents. This way, EGM production teams can centralize and standardize their queries and run them simultaneously with a single call distributed across all three servers.

If all publishing houses, databases, and scientific literature repositories developed similar API implementations, all EGMs could be created with a couple of iterative search queries that returned all matched documents all at once. Researchers would not need to manually search and screen documents on each individual source and queries would not differ from source to source, meaning that more consistency could be attained regarding both matched studies and methodological documentation containing query parameters. This means that the actions outlined above in step iii) of the methodological approach to create EGMs would be greatly simplified and as frictionless as possible.

The other aspect of EGM creation that can be potentially optimized by means of automation relates to step iv) of the methodology. During that process, researchers must manually read, analyze and code all relevant bits of information that will become the EGM. It is, as it sounds, a labor-intensive task that can also be prone to different appraisal outcomes regarding the myriad of important data that has to be extracted from all included documents in the study. This deals essentially with the fact that different people can interpret and therefore decide to include or exclude information in different ways, even if a rigorous methodology to do so has been discussed and put in place in advance. This would render the resulting EGM unreliable, a reason for which a novel approach should be considered and adopted.

Precisely, the approach we propose here tackles the issue of human error and makes the process of data extraction much more efficient. We explored different methods to computationally analyze textual data, particularly Latent Dirichlet Allocation (LDA) models. LDA models for text corpora analysis were introduced in 2003 (Blei et al., 2003) and have had a major impact on computational methods to model the statistical structures present in documents and within document corpora across various disciplines. In a nutshell, LDA is a generative probabilistic approach that models both topic-word and document-topic probability distributions parting from a "bag of words" notion that emanates from an exchangeability assumption, which aim to capture latent variables of abstract notions like topics (Blei et al., 2003). This methodology allows researchers to identify the latent topics that underlie a certain document corpus, which can prove very useful, especially when facing large volumes of documents and no labeled data to approach the task as a supervised classification problem. Even if the task was to be conducted in such a way, a very labor intensive and manual set of steps would need to be taken to substantially assess and code the information contained in the documents (like the information extraction step outlined above for EGMs), which also lacks scalability (Eshima et al., 2020).

LDA models thus calculate and output two main collections of estimations: a set of topics and a list of the proportion of most frequent words that most probably produced them; and the set of documents in the corpus with the corresponding lists of the proportion of topics that most probably generated them. The second set of estimations, that is, the document-topic probabilities, are straightforward to interpret, whereas the same cannot be said from the first set of estimations, whose number can be set by the researchers. In fact, these topic-word estimations have proven to be nonsensical at times, hard to interpret and even share apparently very similar content across different topics (Morstatter & Liu, 2016; Newman et al., 2011). Moreover, the choice of the number of topics to estimate has been shown to have a significant impact on the results obtained (Roberts et al., 2016).

For these reasons, we have developed a semi-automatic tool to cover the EGM construction process from stage ii) to stage v). This tool connects directly to the available APIs of indexed databases of academic and scientific literature to query and download papers that meet the criteria defined in stages ii) and iii) by experts in the field (i.e., the users or policy experts). Then, in the next stage, the user must select and/or discard documents that are not relevant to the research. The next stage includes the NLP estimation, where we use a novel modification of the Latent Dirichlet Allocation (LDA) algorithm called Keyword Assisted Topic Models (keyATM) (Eshima et al., 2020), an extension of a previous model (Jagarlamudi et al., 2012). As the name implies, this model is based on a set of keywords carefully selected by the user to signal the model what keywords belong to which topic according to their specific substantive policy knowledge.

These words are included in the model as topic labels before fitting, therefore eliminating interpretations of dubious topics in later stages and allowing multiple keywords to describe different topics (Eshima et al., 2020). Here, each document that is part of the query is automatically evaluated by the NLP module to find out the probability of one or more interventions (modeled as the topics being identified by keyATM given specialized keywords) within the document. Those with the highest probability of occurrence are presented to the user, who must then evaluate the presence of said intervention (topic) within each of the documents and its corresponding effect, as identified in the documents. As a final step, the tool consolidates the information and generates an interactive visualization of the EGM.

Figure 1 shows the general scheme of the EGM construction tool and the NLP phase as part of this process.

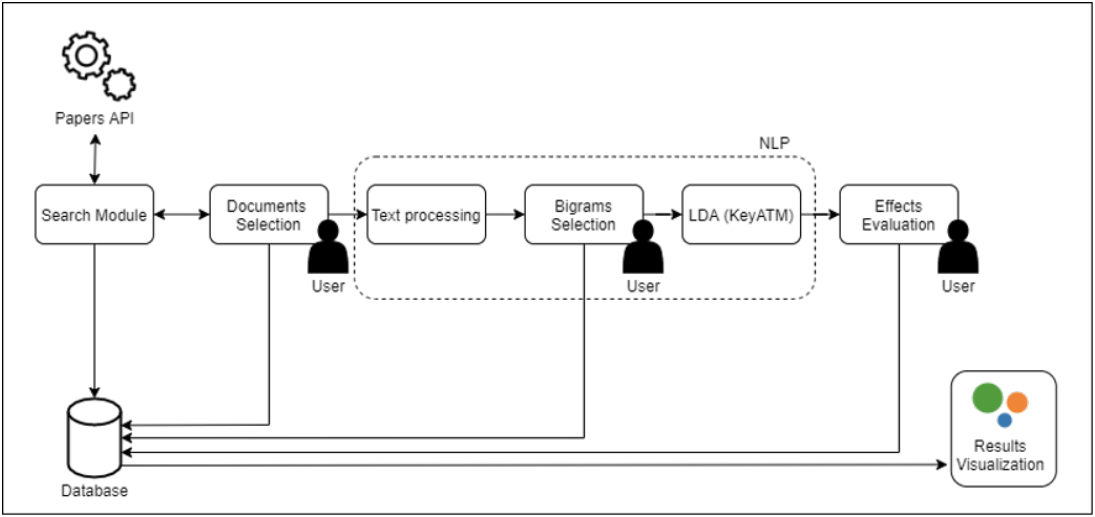

*Figure 1: Block diagram of the tool proposed for the construction of EGM*

This approach is optimal for EGM creation because it essentially allows specialized policy researchers to take advantage of their substantive knowledge and select relevant, content-specific keywords from the corpus and form lists of keywords per topic to the model before fitting it. Also ideal is the fact that the topics that are being identified by the model can also be the corresponding outcome variables of interest determined in the EGM framework, allowing for the document-topic distribution to be employed to signal the policy research team that each document included in the study has been generated from certain topics in each proportion. This means that the researchers would not need to spend as much time manually reading each document to determine when a given outcome is being spoken about, but rather would get into each document with cues from the model regarding what outcomes are studied in each, resulting in the task being significantly reduced to simply corroborating the model's output and identifying what kind of effect has been found for each intervention-outcome pair, whether positive, negative or non-significant.

This approach would not only make the information retrieval and coding much more efficient, but it would also ensure that there is no room for human-driven discrepancies in the information retrieval process, all the while maintaining valuable scalability characteristics and reproducible output.

## 5  Limitations

Although preliminary tests have showed consistent accuracy when applying keyATM to policy-related academic literature for the creation of EGMs, more testing and iteration is needed to improve the model and its proposed use.

This approach has the potential to make EGM creation a much more frictionless, efficient, and timely process. Of course, it does not intend to completely override human interventions in the process. Rather, it seeks to automate specific sub-tasks that would make human interaction more seamless and control-focused, while also retaining crucial tasks, like determining the direction of the effects found for each intervention-outcome pair.

## 6  Discussion and Conclusions

Unquestionably, the COVID-19 pandemic has left governments facing major crises and challenges, but also with some advantages, in terms of information and extensive evidence, very useful for improving the design, formulation, implementation and evaluation of public policies.

In this sense, this document presented a methodological review of EGMs and the development of an information gathering and processing system to facilitate the EGM construction process, from the determination of inclusion criteria to the analysis and user-friendly presentation.

Our aim is to show how our proposed system can further enable policy makers around the world to have efficient and timely access to all the evidence they need to make decisions, while at the same time communicating vast amount of information in a visual and friendly manner.

Although still in early iteration stages, we believe this approach can facilitate the creation of EGMs, namely by substantially reducing the time it takes to create them; avoiding bias; allowing for more systematized, robust, and replicable methodologies and information management; avoiding human error; enabling access to potentially larger universes of evidence; among many others not yet researched affordances.

15